\documentclass[apjl]{emulateapj}
\newcommand {\gsim}{ \lower .75ex \hbox{$\sim$} \llap{\raise .27ex \hbox{$>$}} } 
\newcommand {\lsim}{ \lower .75ex\hbox{$\sim$} \llap{\raise .27ex \hbox{$<$}} }

\newcommand {\sw}{{\it Swift}}

\newcommand {\fe}{{\it Fermi}}

\shorttitle{The GeV afterglow of GRB090510}
\shortauthors{Ghirlanda et al.}

\begin{document}

\title{The onset of the GeV afterglow of GRB 090510}

\author{G. Ghirlanda\altaffilmark{1}, G. Ghisellini\altaffilmark{1} and  L. Nava\altaffilmark{1,2}
}
%\affil{INAF--Osservatorio Astronomico di Brera, via Bianchi 46, I--23807 Merate, Italy}
\email{giancarlo.ghirlanda@brera.inaf.it}
%\and
%\author{L. Nava\altaffilmark{1,2}}
%\affil{Univ. dell'Insubria, V. Valleggio, 11, I--22100, Como, Italy}


\altaffiltext{1}{INAF--Osservatorio Astronomico di Brera, via Bianchi 46, I--23807 Merate, Italy}
\altaffiltext{2}{Univ. dell'Insubria, V. Valleggio, 11, I--22100, Como, Italy}

\begin{abstract}

We study the emission of the short/hard GRB 090510 at energies $>0.1$ GeV 
as observed by the Large Area Telescope (LAT) onboard the \fe\ satellite.
The GeV flux rises in time as $t^2$ and decays as $t^{-1.5}$ up to 200 s.
The peak of the high energy flux is delayed by 0.2 s with respect to 
the main $\sim$MeV pulse detected by the \fe\ Gamma Burst Monitor (GBM).  
Its energy spectrum is consistent with $F(\nu)\propto \nu^{-1}$.
The time behavior and the spectrum of the high energy LAT flux 
are strong evidences of an afterglow origin.
We then interpret it as synchrotron radiation produced by the forward shock 
of a fireball having a bulk Lorentz factor $\Gamma \sim 2000$. 
The afterglow peak time is independent of energy in the 0.1--30 GeV range
and coincides with the arrival time of the highest energy photon ($\sim$ 30 GeV).
Since the flux detected by the GBM and the LAT have
different origins, the delay between these two components 
is not entirely due to possible violation of the Lorentz invariance.
It is the LAT component by itself that allows to set a stringent lower limit 
on the quantum--gravity mass of 4.7 times the Planck mass. 
\end{abstract}


\keywords{Gamma rays: bursts --- Radiation mechanisms: non-thermal --- X--rays: general}

\section{Introduction}

%\section{GRB 090510}

GRB 090510 is a short/hard burst at redshift $z$=0.903$\pm$0.003 (Rau et al. 2009)  
detected by \fe\ (Guiriec et al. 2009), AGILE (Longo et al. 2009), \sw\ (Hoversten et al. 2009), 
{\it Konus-Wind} (Golenetskii et al. 2009) and {\it Suzaku} (Ohmori et al. 2009).

The \fe--Gamma Burst Monitor (GBM) triggered on a precursor while 
the main emission episode in the 8 keV--40 MeV energy range 
starts $\sim$0.5 s after trigger and lasts up to $\sim$1 s. 
The emission observed by the \fe--Large Area Telescope (LAT) starts 0.65 s after the trigger 
and lasts $\sim$ 200 s.  
The joint GBM--LAT spectral analysis showed 
the presence of two components. 
\fe--LAT detected a 31$\pm3$ GeV photon delayed by 0.829 s with respect 
to the trigger (Abdo et al. 2009 -- A09 hereafter). 

The precursor was not seen by AGILE, that triggered on  the main emission episode.  
The flux detected by the Mini Calorimeter (MCAL, 0.3--10 MeV) lasts 0.2 s. 
As it ends, the Gamma Ray Imaging Detector (GRID, 0.03 -- 30 GeV) 
starts to detect a high energy component lasting for 10 s and decaying as  
$t^{-1.3}$ (Giuliani et al. 2009).

\begin{figure*}[hbt]
\centering
\includegraphics[scale=0.9]{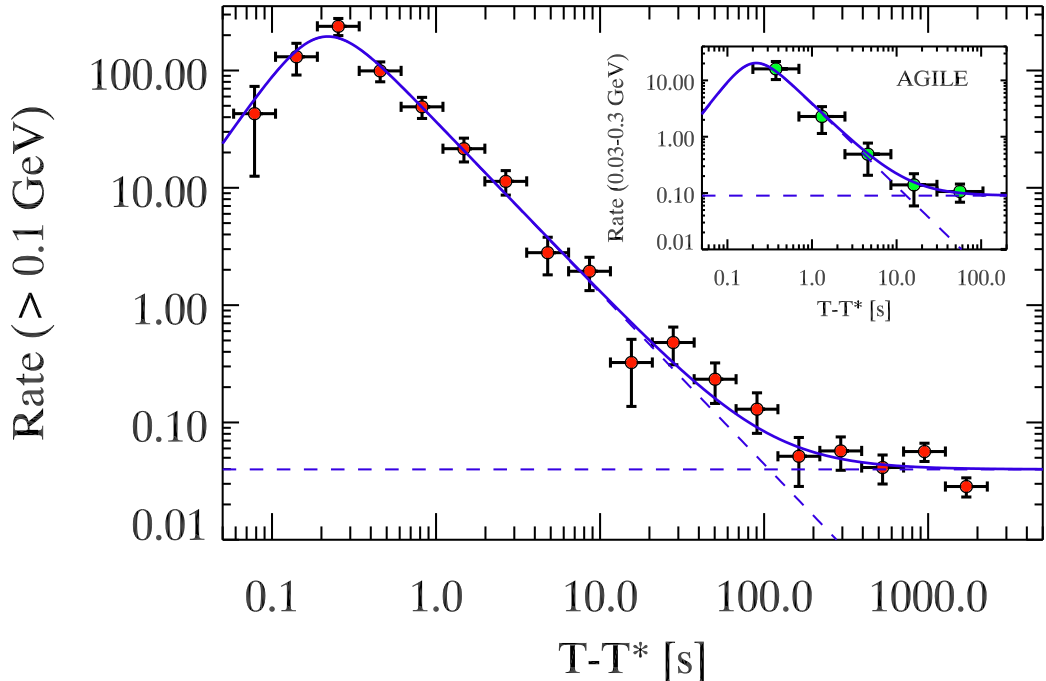}
\vskip 0.5cm
\caption{
\fe--LAT light curve of the emission of GRB 090510 above 100 MeV. 
The times were scaled to the time $T^{*}=0.6$ s after the GBM trigger. 
This is the time corresponding to the main pulse of
emission detected by the GBM in the 8 keV--10 MeV energy range (see A09). 
The solid line is the best fit to the data points obtained with a smoothly 
broken power--law plus a constant (dashed lines). 
The inset shows the AGILE light curve (photons energies between 
30 and 300 MeV). 
The curve is the best fit of the \fe\ data scaled to the 
AGILE data points.
}
\label{fg1}
\end{figure*}

The distinguishing property of GRB 090510 arising from the \fe\ and AGILE data 
is that the $\sim$MeV emission component, commonly detected in GRBs, is followed 
by a much longer lasting high energy emission detected above 100 MeV. 
Both the AGILE and \fe\ spectra suggest that this component is not 
the extrapolation of the soft $\sim$MeV spectrum to the GeV range. 
A09 interpret the $\sim$MeV flux as synchrotron radiation and the 
LAT flux as its synchrotron self--Compton emission. 
The detection by \fe\ of a 30 GeV photon sets a lower limit on the bulk Lorentz 
factor of the fireball $\Gamma>1000$, based on the compactness argument (A09). 
The 30 GeV photon arrives 0.829 s after the trigger (set by the precursor) and 
0.3 s after the beginning of the GBM main pulse. These delays allowed A09 to put 
limits on the violation of the Lorentz invariance. 

In this paper we propose a different interpretation of the emission detected by
the \fe--LAT and AGILE.
If $\Gamma>1000$ the fireball should start to decelerate and produce  
a luminous afterglow rather early (e.g. Piran 2005), even at the sub--second timescale.
By analyzing the \fe--LAT light--curve and spectra
we present strong evidences that the flux detected by the LAT
is afterglow emission of the forward external shock.
%
% that the flux rises as $t^2$, peaks at $t=T-T^{*}$=0.2 s and then declines as 
%$t^{-1.5}$ ($T^{*}=0.6$s corresponds to the time of the main pulse detected by the GBM).  
%This is a strong evidence that the LAT flux 
%is due to afterglow emission of the forward external shock. 
%
In this framework we derive the initial $\Gamma$ of the fireball 
and set a lower limit on the quantum--gravity mass. 

Recent works on the high energy emission of LAT--detected GRBs
include Kumar \& Barniol Duran (2009), discussing GRB 080916C 
(Abdo et al. 2009a).
Also for this burst they proposed that the LAT--detected flux can be
synchrotron produced in the external shock (see also Gao et al. 2009 for GRB 090510;
see also Fan et al. 2008; Zou, Fan \& Piran 2009; Zhang \& Peer 2009 
for an inverse Compton origin). 
Hadronic models have been proposed to explain the emission properties of 
GRB 080916C (Razzaque, Dermer \& Finke 2009).

\section{\fe--LAT data analysis}

We have analyzed the \fe--LAT data of GRB 090510 with the \fe\ 
\texttt{ScienceTools} \texttt{(v9r15p2)} released on Aug. 8th 2008. 
Photons were selected (with the \texttt{gtselect} tool) around 
RA=333.552$^{\circ}$ and Dec=--26.598$^{\circ}$.  
Different energy bins were considered for the analysis of the LAT light 
curve but only photons with energy $>$100 MeV were extracted. 
Light curves and spectra were created with the \texttt{gtbin} tool. 
The spectral response files were created with the \texttt{gtrspgen}.  
The spectra were analyzed with \texttt{Xspec(v.12)}.

\section{Results}

Fig. \ref{fg1}  shows the light curve considering all the LAT photons with energies $>$100 MeV. 
The times are scaled to $T^{*}=0.6$ s which corresponds to the time of the first main pulse 
observed by the GBM. 
We fit the light curve with the sum of two components, i.e. a 
smoothly broken power--law and a constant to account 
for the flattening of the flux visible after 200 s:
\begin{equation}
R(t)= {{{A\, (t/t_{\rm b}})^{\alpha}}\over{1+({t/t_{\rm b}})^{\alpha+\beta}}}+B
\label{f1}
\end{equation} 
When $\alpha>0$ and $\beta>0$ Eq. \ref{f1} has a peak at 
$t_{\rm peak}=t_{\rm b}\,(\alpha/\beta)^{1/(\alpha+\beta)}$.
The standard afterglow theory (Sari \& Piran 1999) requires $\alpha=2$.
Fixing $\alpha=2$, 
the best fit parameters ($\chi^2$=14.6/14) are $A=385^{+45}_{-40}$ counts/s, 
$t_{\rm pk}=0.217\pm0.015$ s, $\beta=1.46^{+0.06}_{-0.03}$ and $B=4\times 10^{-2}$ counts/s. 
The best fit is shown by the solid line in Fig. \ref{fg1}. 
The AGILE light curve (adapted from Giuliani et al. 2009) of the photons detected by 
the GRID between 30 MeV and 300 MeV is also shown in Fig. \ref{fg1} (inset).
The \fe\ and AGILE light curves are consistent with the same decay law, i.e. $t^{-1.5}$, 
although AGILE missed the rising phase of the GeV emission. 

\begin{figure}[hbt]
\centering
\includegraphics[scale=.50]{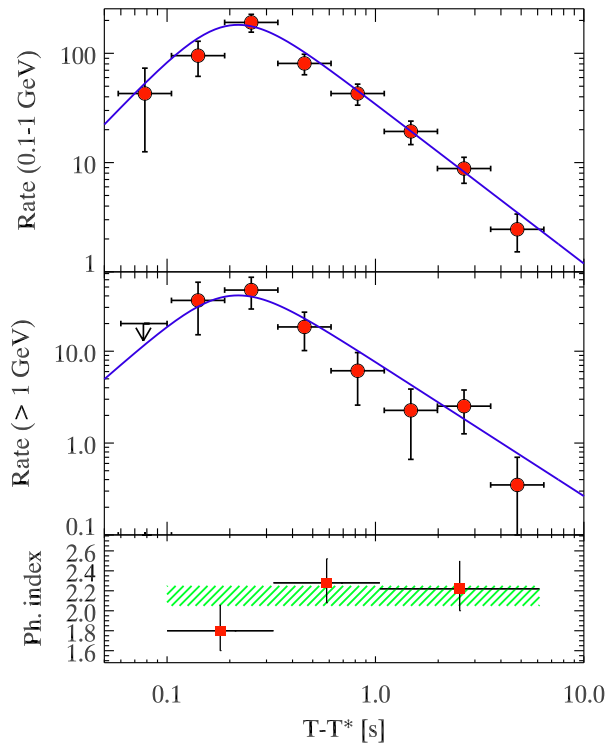}
\caption{
\fe--LAT light curve of GRB 090510 between 0.1 and 1 GeV 
and above 1 GeV
(top and middle panels, respectively) in the first 10 seconds. 
The times are scaled to $T^{*}$=0.6 s (see text). 
The solid line is the fit of the light curve $>$0.1 GeV (Fig. \ref{fg1}). 
The bottom panel shows the photon spectral index (1$\sigma$ errors are shown) 
of the LAT spectra for the time--integrated spectrum (hatched region) and for 
three time resolved spectra (squares).  
}
\label{fg2}
\end{figure}

The emission above 100 MeV peaks at $T-T^{*}$=0.22 s (i.e. 0.82 s after the GBM trigger). 
The time of the peak coincides with the arrival time 
of the highest energy photon of 30 GeV. % detected by the LAT from GRB090510. 
Fig. \ref{fg1} shows that the LAT flux lasts for about 200 s 
(and it sets to the background level afterwords). 
Instead, the emission detected 
by the GBM in the 8 keV--10 MeV energy range ceases after $\sim$1 s (A09). 
% The GeV light curve rises as $t^2$ and decreases as $t^{-1.5}$ before and after $t_{\rm pk}$.

\begin{figure}[!h]
\centering
\includegraphics[scale=.52]{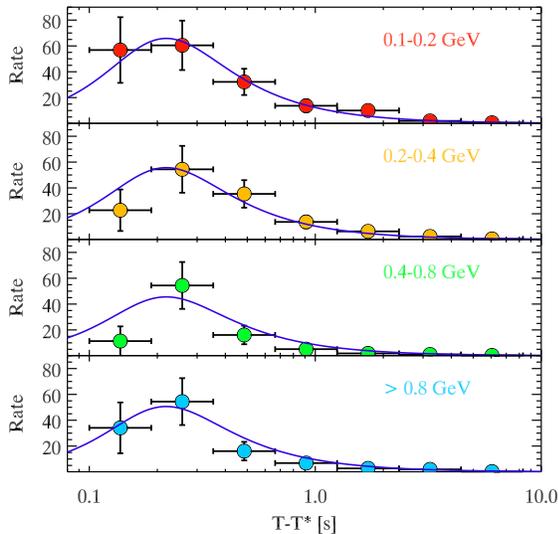}
\caption{
\fe--LAT light curve of GRB 090510 in four energy channels 
(from top to bottom): 0.1--0.2 GeV, 0.2--0.4 GeV, 0.4--0.8 GeV, $>$0.8 GeV. 
The curves are the best fit obtained from the LAT light curve ($>$0.1 GeV - Fig. \ref{fg1}) 
rescaled to the single channel light curves.     
}
\label{fg3}
\end{figure}

Fig. \ref{fg2} shows the LAT light curve in the first 10 s separated in two energy bands, 
i.e. 0.1--1 GeV and $>$1 GeV (top and middle panels, respectively). 
The curves correspond to the same best fit obtained from the $>$0.1 GeV light curve 
(Fig. \ref{fg1}), only re--normalized to the data points. 
We further separated the light curve into four broad energy channels: 
0.1--0.2 GeV, 0.2--0.4 GeV, 0.4--0.8 GeV and $>$0.8 GeV (Fig. \ref{fg3}). 
Fig. \ref{fg2} and Fig. \ref{fg3} show that the time of the peak is 
the same in different energy ranges.  

We also analyzed the spectra of the early GeV emission component. 
We considered the spectrum integrated in time between $T-T^{*}$=0.1 and 7 s 
and we also extracted three time resolved spectra distributed in this time interval. 
The photon spectral index of the fit with a single power 
law of the average spectrum (hatched region) and of the time resolved 
spectra (filled squares) are shown in the bottom panel of Fig. \ref{fg2}. 
The spectrum before the peak is hard with a photon index 1.87$\pm$0.2 and then 
it softens to 2.2$\pm$0.2.
Both are consistent with the time--integrated spectrum.

\section{Estimate of the initial bulk Lorentz factor}

The derived peak time of the LAT received flux translates
into an estimate of the bulk Lorentz factor $\Gamma_0$ at the start
of the afterglow.
The peak time of the afterglow bolometric luminosity occurs at a
time of the order of the deceleration time.
If the circumburst number density $n$ is homogeneous we have (e.g. Sari \& Piran 1999):
\begin{equation}
{ t_{\rm peak} \over 1+z} \, \sim \, t_{\rm dec} \, \sim \,
\left( { 3E_{\rm k,iso} \over
32 \pi n m_{\rm p} c^5 \Gamma_0^8} \right)^{1/3}
\label{tpeak}
\end{equation}
where $E_{\rm k,iso}$ is the isotropic kinetic energy of the fireball,
estimated through the emitted energy of the prompt emission assuming
an efficiency $\eta$ ($E_{\rm k,iso}=E_{\rm \gamma,iso}/\eta$).
% and $n$ is the number density of protons of the circumburst medium.
We use Eq. \ref{tpeak} to estimate $\Gamma_0$.
Setting $E_{\rm \gamma,iso}=3.5\times 10^{52}$ erg (A09, excluding the
LAT component), $\eta=0.2$ and  $t_{\rm peak}=0.2$ s we derive
$\Gamma_0=1.96 \times 10^3 \, n^{-1/8}$.
This value is not much larger than the lower limits derived by A09
through the compactness argument and assuming that the LAT component
belongs to the prompt emission.
It is also rather insensitive to the (unknown) particle density $n$.
The distance from the central engine corresponding to the peak time is
$R_{\rm peak} \sim 2 c t_{\rm peak}  \Gamma_0^2/(1+z)= 2.4\times
10^{16}n^{-1/4}$ cm.

\subsection{A synchrotron origin of the LAT emission}

Following standard arguments, the minimum electron energy
of the injected electrons in the forward shock is
$\gamma_{\rm m} \sim \epsilon_{\rm e} \Gamma m_{\rm p}/m_{\rm e}$,
while the magnetic field value is
$B\sim \Gamma (8\pi \epsilon_{\rm B} n m_{\rm p}c^2)^{1/2}$.
Electrons with $\gamma_{\rm m}$ emit an observed synchrotron frequency
$\nu_{\rm m} \sim 2\Gamma (4/3) eB/(2\pi m_{\rm e} c) \gamma_{\rm
m}^2/(1+z)\sim
10.6\, \Gamma_3^4 (n \epsilon_{\rm B} )^{1/2} \epsilon_{\rm e}^2/(1+z)$ MeV.
This frequency is below the LAT energy range, but the injection of a
power law
distribution of electrons extending to $\gamma_{\rm max}\sim
(10^2$--$10^3)\gamma_{\rm m}$
ensures that the LAT flux can indeed have a synchrotron origin.
The synchrotron self--Compton (SSC) spectrum extends to much higher
frequencies (e.g. Fan et al. 2008, Corsi et al. 2009),
but becomes important only above $\nu_{\rm m,C}\sim \gamma_{\rm m}^2
\nu_{\rm m}$,
i.e. above the TeV energy range.
Note the strong dependence of $\nu_{\rm m}$ and $\nu_{\rm m,C}$ on the bulk
Lorentz factor:
a synchrotron origin of a $\sim$GeV afterglow is reasonable only for
rather large $\Gamma_0$,
while the SSC flux becomes more important for smaller $\Gamma_0$.

This has a simple and important consequence. Bursts with $\Gamma_0\sim 100$ 
or smaller can produce high energy afterglow radiation through the SSC mechanism 
but {\it the onset time of their afterglows} will be large, in turn implying, for the same 
emitted energy, a lower luminosity (see also Kumar \& Barniol Duran 2009). 
They are then more difficult to detect. The best candidates for a LAT detection 
are therefore bursts with large $\Gamma_0$, because this ensures an early onset 
of the afterglow, implying large luminosities.

\section{Test of the Lorentz--invariance violation}

The arrival time of the 30 GeV photon coincides with the peak of the afterglow emission. 
This is reasonable, because this is the time when we have the maximum probability to detect it 
(maximum flux and hard spectrum). 
{\it If we assume} that the 30 GeV photon was indeed produced at the 
peak time, then the maximum possible time delay it can have is 
of the order of the width of the time bin of the peak (0.15 s).
More conservatively, we can assume that the 30 GeV photon was produced right at the 
beginning of the afterglow and it arrives delayed by 0.22 s due to violation of the 
Lorentz invariance.

The time delay $\Delta t$ between the arrival time of a low and a high energy photon 
in the case of a linear dependence of the photon's propagation speed on its energy is 
\begin{equation}
\Delta t\, =\,  { {{\Delta E}\over{M_{\rm QG} c^2}} \, 
\int_{0}^{z}  {{1+z} 
\over { H_{0} \sqrt{ \Omega_{m} (1+z)^{3} + \Omega_{\Lambda} } }}\, dz }
\end{equation}
where $\Delta E$ is the difference between the low and high photon energy and 
$M_{\rm QG}$ is the quantum--gravity mass (A09; Amelino--Camelia et al. 1998;
Jacob \& Piran 2008). 

For a delay of 0.15 s we derive  $M_{\rm QG} > 6.7 M_{\rm Planck}$ while the more 
conservative limit (delay of 0.217 s) is $M_{\rm QG} > 4.7 M_{\rm Planck}$
(we used $h_0=0.71, \Omega_\Lambda=0.73, \Omega_{\rm M}=0.27$).
These limits are consistent with those of A09, but excludes their lowest estimates.

\section{Conclusions}

The detection of an early high energy emission in the GeV range
inevitably flags a large bulk Lorentz factor $\Gamma_0$: to avoid
suppression of the GeV emission due to the $\gamma$--$\gamma \to e^\pm$ process
if the high energy photons belong to the prompt phase, or to have
an early peak flux time if the emission belongs to the afterglow phase.

We have shown that the latter case is indeed favored (see also Gao et al. 2009), 
because we see the peak time of the emission in the LAT light--curve
(as also seen in other GRBs in the infrared--optical band, see e.g.
Molinari et al. 2007 for GRB 060418 and GRB 060607A).
Furthermore, also the energy spectral index $F(\nu)\propto \nu^{-1}$
is very similar to the ones we see in the afterglow phase.
A large $\Gamma_0$, implying an early onset of the afterglow, means a large
luminosity at the peak time (for equal emitted energy), and large
typical frequencies. This makes synchrotron the most likely process for the 
LAT emission we see.

GRBs with smaller $\Gamma_0$ will have their prompt emission less blue--shifted,
and it would be more difficult for them to reach the LAT energy range
during their prompt phase. 
Their afterglows can, through the SSC process,
but their afterglow peak time is longer, and so their fluxes are fainter
(as $t_{\rm peak}^{-1} \propto \Gamma_0^{8/3}$ if they emit, at the peak,
the same amount of energy, see Eq. 2).
A large $\Gamma_0$, instead, means a large blue--shift for the photons of the prompt,
an early onset of the afterglow, implying more flux at the peak, and finally
larger intrinsic afterglow frequencies, allowing even the synchrotron 
photons of the afterglow to reach the LAT energy range.
Therefore GRBs with large $\Gamma_0$ should be much more luminous in the
LAT energy range than the other GRBs (see also Kumar \& Barniol Duran 2009).

The limits derived here on the quantum gravity mass scale are not very
different from the ones derived by A09, but we could
associate the GBM and the LAT fluxes to two different components.
We can then argue that
the high energy photons are {\it generated} at (slightly) later 
times than the photons detected by the GBM, and the delay of their {\it arrival}
times is not entirely due to quantum gravity effects.
Instead, since photons above 100 MeV belong to the same
component they are the best tool to investigate
quantum gravity effects.

This suggests a recipe for a robust test on the Lorentz invariance
violation, possible with very {\it bright} and {\it short} bursts detected 
at high energies.
A {\it short} duration of the prompt ensures that the fireball has
a relatively narrow width, and in turn this should correspond
to a well--defined afterglow peak.
A {\it bright} flux ensures good photon statistics, enabling to measure more
accurately possible delays as a function of photon energies.

\acknowledgments
This research was supported by PRIN--INAF 2009 and ASI
I/088/06/ grants. G. Ghirlanda acknowledges the NORDITA program 
on Physics of relativistic flows. 
We thank F. Tavecchio and Y. Poutanen for helpful discussions.

%% Here we use \plottwo to present two versions of the same figure,
%% one in black and white for print the other in RGB color
%% for online presentation. Note that the caption indicates
%% that a color version of the figure will be available online.
%%

\end{document}